\documentclass[iop,apjl]{emulateapj}

\newcommand{\simless}{\mathbin{\lower 3pt\hbox
      {$\rlap{\raise 5pt\hbox{$\char'074$}}\mathchar"7218$}}} 
\newcommand{\simgreat}{\mathbin{\lower 3pt\hbox
     {$\rlap{\raise 5pt\hbox{$\char'076$}}\mathchar"7218$}}} 

\slugcomment{\today}

\shorttitle{ALMA observations of 2M1207}
\shortauthors{Ricci et al.}

\begin{document}


\title{ALMA Observations of the Young Substellar Binary System 2M1207}


\author{L. Ricci}

\affil{Department of Physics and Astronomy, Rice University, 6100 Main Street, 77005 Houston, TX, USA}
\affil{Harvard-Smithsonian Center for Astrophysics, 60 Garden Street, 02138 Cambridge, MA, USA}

\and

\author{P. Cazzoletti}
\affil{Max-Planck-Institut f\"{u}r
extraterrestrische Physik, Giessenbachstrasse 1, 85748 Garching, Germany}

\and

\author{I. Czekala, S. M. Andrews, D. Wilner}
\affil{Harvard-Smithsonian Center for Astrophysics, 60 Garden Street, 02138 Cambridge, MA, USA}

\and

\author{L. Sz\H{u}cs}
\affil{Max-Planck-Institut f\"{u}r
extraterrestrische Physik, Giessenbachstrasse 1, 85748 Garching, Germany}

\and

\author{G. Lodato}
\affil{Dipartimento di Fisica, Universit\'aˆ degli studi di Milano, Via Celoria 16, 20133 Milano, Italy}

\and

\author{L. Testi}
\affil{European Southern Observatory (ESO) Headquarters, Karl-Schwarzschild-Str. 2, 85748 Garching, Germany}

\and

\author{I. Pascucci}
\affil{Lunar and Planetary Laboratory, University of Arizona, 85721 Tucson, AZ, USA}

\and

\author{S. Mohanty}
\affil{Imperial College London,  1010 Blackett Lab.,  Prince Consort Road, London SW7 2AZ, UK}

\and

\author{D. Apai}
\affil{Lunar and Planetary Laboratory and Steward Observatory, University of Arizona, 85721 Tucson, AZ, USA}

\and

\author{J. M. Carpenter}
\affil{Joint ALMA Observatory, Alonso de Córdova 3107, 763 0355 Vitacura, Santiago, Chile}

\and

\author{B. P. Bowler}
\affil{McDonald Observatory and University of Texas, Department of Astronomy, 2515 Speedway, 78712, Austin, TX, USA}
\affil{Hubble Fellow}

\email{luca.ricci@cfa.harvard.edu}


\begin{abstract}

We present ALMA observations of the 2M1207 system, a young binary made of a brown dwarf with a planetary-mass companion at a projected separation of about 40 au.
We detect emission from dust continuum at 0.89 mm and from the  $J = 3 - 2$ rotational transition of CO from a very compact disk around the young brown dwarf.
The small radius found for this brown dwarf disk may be due to truncation from the tidal interaction with the planetary-mass companion.  Under the assumption of optically thin dust emission, we estimated a dust mass of 0.1 $M_{\oplus}$ for the 2M1207A disk, and a 3$\sigma$ upper limit of $\sim 1~M_{\rm{Moon}}$ for dust surrounding 2M1207b, which is the tightest upper limit obtained so far for the mass of dust particles surrounding a young planetary-mass companion.
We discuss the impact of this and other non-detections of young planetary-mass companions for models of planet formation, which predict the presence of circum-planetary material surrounding these objects.
    

\end{abstract}



\section{Introduction}
\label{sec:intro}

Planets form out of the solids and gas present in young circumstellar disks. The past two decades have witnessed giant leaps in the characterization of the demographics, structure and evolution of disks surrounding young stars \citep[see][for a recent review]{Andrews:2015}.

Disks have been routinely found also around young brown dwarfs, objects intermediate in mass between stars and planets \citep{Comeron:1998,Natta:2001,Muench:2001,Klein:2003,Bayo:2017}. The study and characterization of brown dwarf disks is particularly relevant to investigate the potential of finding exoplanets around more evolved brown dwarfs \citep{Payne:2007,Ricci:2014}, as well as to test theories of disk evolution and planet formation under physical conditions which can be very different from the ones probed in disks surrounding young stars \citep{Pinilla:2013, Meru:2013}.

More than a hundred brown dwarf disks have been detected in the infrared \citep[e.g.][]{Jayawardhana:2003,Liu:2003,Scholz:2007,Harvey:2012a,Harvey:2012b}. 
Compared with these observations, interferometric observations in the sub-millimeter/millimeter can obtain better angular resolution, and directly constrain the spatial distribution of dust and gas in the disk.


Here we present Atacama Large Millimeter Array (ALMA) observations of the 2MASS J12073346-3932539 (henceforth 2M1207) system, made of a young brown dwarf \citep[2M1207A, M8-spectral type, $M_{\rm{1207A}} \sim 25~M_{\rm{Jup}}$,][]{Mohanty:2007} and the first directly imaged extra-solar planetary-mass companion \citep[2M1207b, $M_{\rm{1207b}} \sim 5 \pm 2 ~ M_{\rm{Jup}}$,][]{Chauvin:2004,Chauvin:2005,Song:2006,Bowler:2016}. The 2M1207 system is a likely member of the TW Hya association, with an estimated age of $\approx 5 - 10$ Myr~\citep{Gizis:2002,Mamajek:2005,Weinberger:2012}. 
2M1207A and b have an angular separation of 0.77$''$, which corresponds to a projected separation of 40.6 au at a distance of 52.8 pc \citep{Ducourant:2008}.
Given the large companion-to-host mass ratio and large separation, \citet{Lodato:2005} proposed gravitational fragmentation of the 2M1207A disk \citep{Sterzik:2004, Riaz:2012} as the most likely formation mechanism for 2M1207b. 

The luminosity of 2M1207b, as derived from near-infrared photometry, is about 2.5 mag lower than that predicted based on its mid- to late L spectral type and effective temperature of $\sim 1600$ K, implying an unphysically small radius to reproduce its spectrum~\citep{Mohanty:2007,Patience:2010}. \citet{Mohanty:2007} proposed dust absorption from an edge-on circum-planetary disk as a possible explanation of 2M1207b's under-luminosity. \citet{Skemer:2011} argued against the edge-on disk hypothesis because of the lack of photometric variability, a fine tuned disk geometry needed to
reproduce the observed spectral energy distribution (SED), and the relatively large sample of under-luminous brown dwarfs/giant exo-planets. They suggested an atmosphere with thick clouds to reproduce the 2M1207b SED. The favored interpretation is that the \textit{apparent} optical and near-infrared under-luminosity is an effect of the redistribution at longer wavelengths of the light from a $T_{\rm{eff}} \sim 1100$ K atmosphere because of thick photospheric dust \citep{Skemer:2011,Barman:2011} . Rotational modulations with a short period of $\approx 10$ hours from the 2M1207b atmosphere were recently detected through high-contrast, high-cadence observations with the \textit{Hubble Space Telescope} \citep{Zhou:2016}.

In all formation scenarios, planetary-mass objects accrete most of their mass from a circum-planetary disk \citep{Vorobyov:2010,Ward:2010}. According to some theoretical studies, the cores of gas-giant planets grow predominantly through a gradual accumulation of pebbles, i.e. mm/cm-sized dust grains, which are efficient emitters in the sub-mm \citep[e.g.][]{Levison:2015}. Hence, significant emission at these wavelengths is expected during, or shortly after, the process of massive planet formation \citep{Zhu:2016}.   

Recent deep observations at sub-mm/mm wavelengths have attempted the detection of circum-planetary disks orbiting candidate planetary-mass objects directly seen in the optical and near infrared \citep[NIR,][]{Isella:2014,Bowler:2015,MacGregor:2016}. 
The angular resolution of our ALMA observations allows us to separate the 2M1207A brown dwarf disk from the emission from possible material surrounding the 2M1207b planetary-mass object. 
Given the sensitivity of the ALMA observations, and the proximity of the 2M1207 system, a factor of $\approx 3$ closer than the younger closest star forming regions, we can investigate the immediate environments of a brown dwarf $-$ planetary mass system at unprecedented depth in terms of dust mass.

Section~\ref{sec:obs} presents the ALMA observations and data reduction. The results of these observations are outlined in Section~\ref{sec:results}. The analysis of the dust continuum and CO data are reported in Section~\ref{sec:continuum_analysis} and~\ref{sec:analysis}, respectively. Section~\ref{sec:discussion} presents the discussion of the results. Conclusions are described in Section~\ref{sec:conclusions}.



\section{ALMA Observations and Data Reduction}
\label{sec:obs}

We observed the 2M1207 system using ALMA Early Science in Cycle 2 at Band 7 (frequency of about 338 GHz). 
Observations were performed on June 29, July 17 and 18 2014, when 30 antennas were available. Baseline lengths ranged between 19.6 and 650.3 m.

The total time on-source was approximately 2 hours. The ALMA correlator was configured to record
dual polarization with four separate spectral windows, each with a total bandwidth of 1.875 GHz. Their central frequency is 330.98, 332.93, 342.98 and 344.98 GHz, respectively. The spectral frequencies of the last and first spectral windows were chosen in order to probe possible molecular emission from $^{12}$CO($J = 3 - 2$) and $^{13}$CO($J = 3 - 2$), respectively. However the  $^{13}$CO($J = 3 - 2$) line was not detected and it is not discussed here. The channel width in these two spectral windows is 0.488 MHz (0.42 km s$^{-1}$ for CO $J = 3 - 2$). The spectral resolution is twice the channel spacing since the data are Hanning smoothed. The mean frequency of the observations is 338.0 GHz ($\lambda = 0.89$ mm).

The ALMA data were calibrated by National Radio Astronomy Observatory (NRAO) staff using the CASA software package version 4.2 \citep{McMullin:2007}.
Simultaneous observations of the 183 GHz water line with the water vapor radiometers were used to reduce atmospheric phase noise before using J1147$-$3812 for standard complex calibration. The same source was used to calibrate the frequency-dependent bandpass. The flux scale was determined with observations of Titan, and adopting the Butler-JPL-Horizon 2012 models, resulting in an accuracy of about 10$\%$.

\begin{figure*}[t!]
\centering
\begin{tabular}{cl}
\includegraphics[scale=0.55]{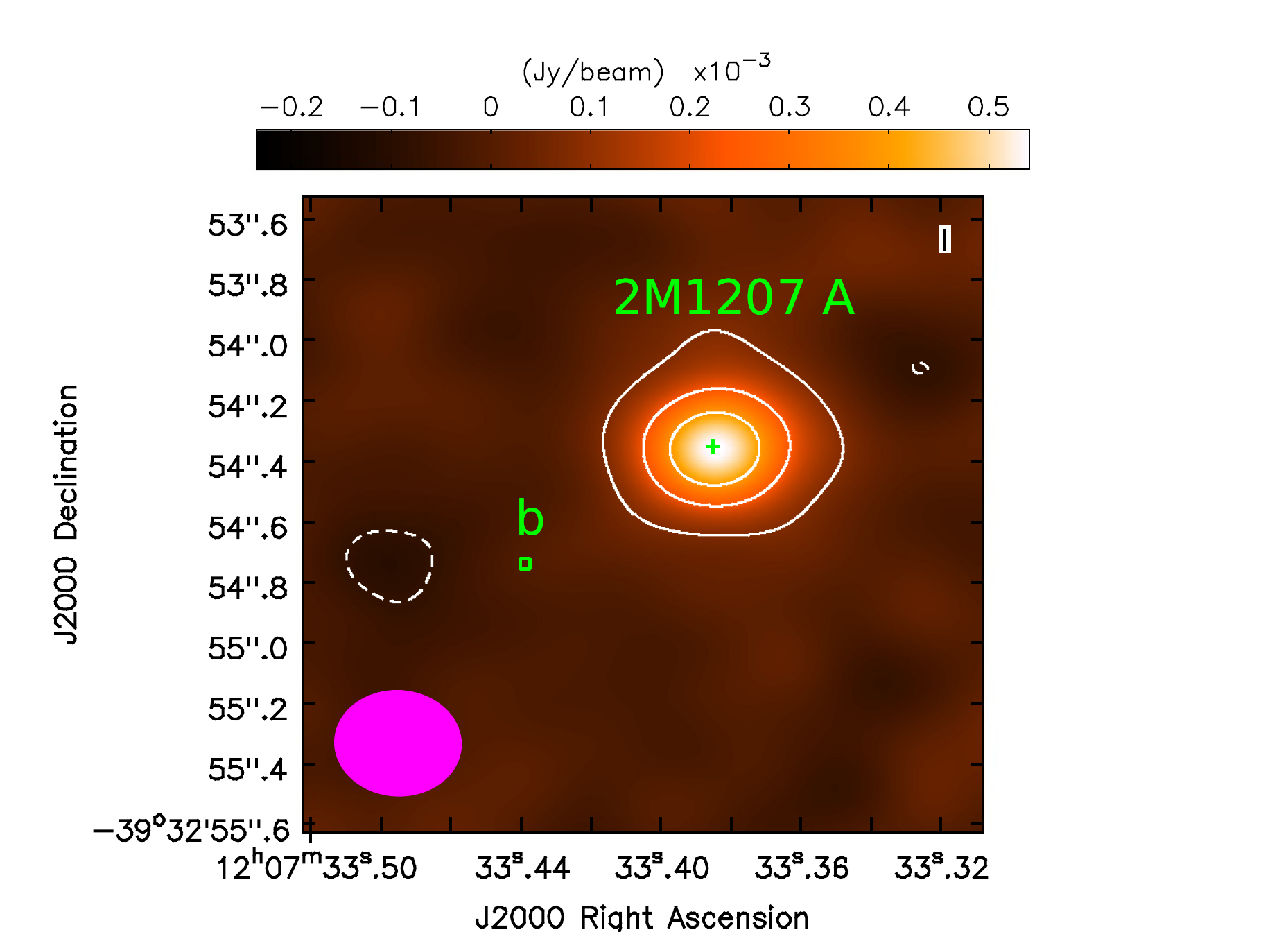}   & \hspace{-25mm}
\includegraphics[scale=0.67]{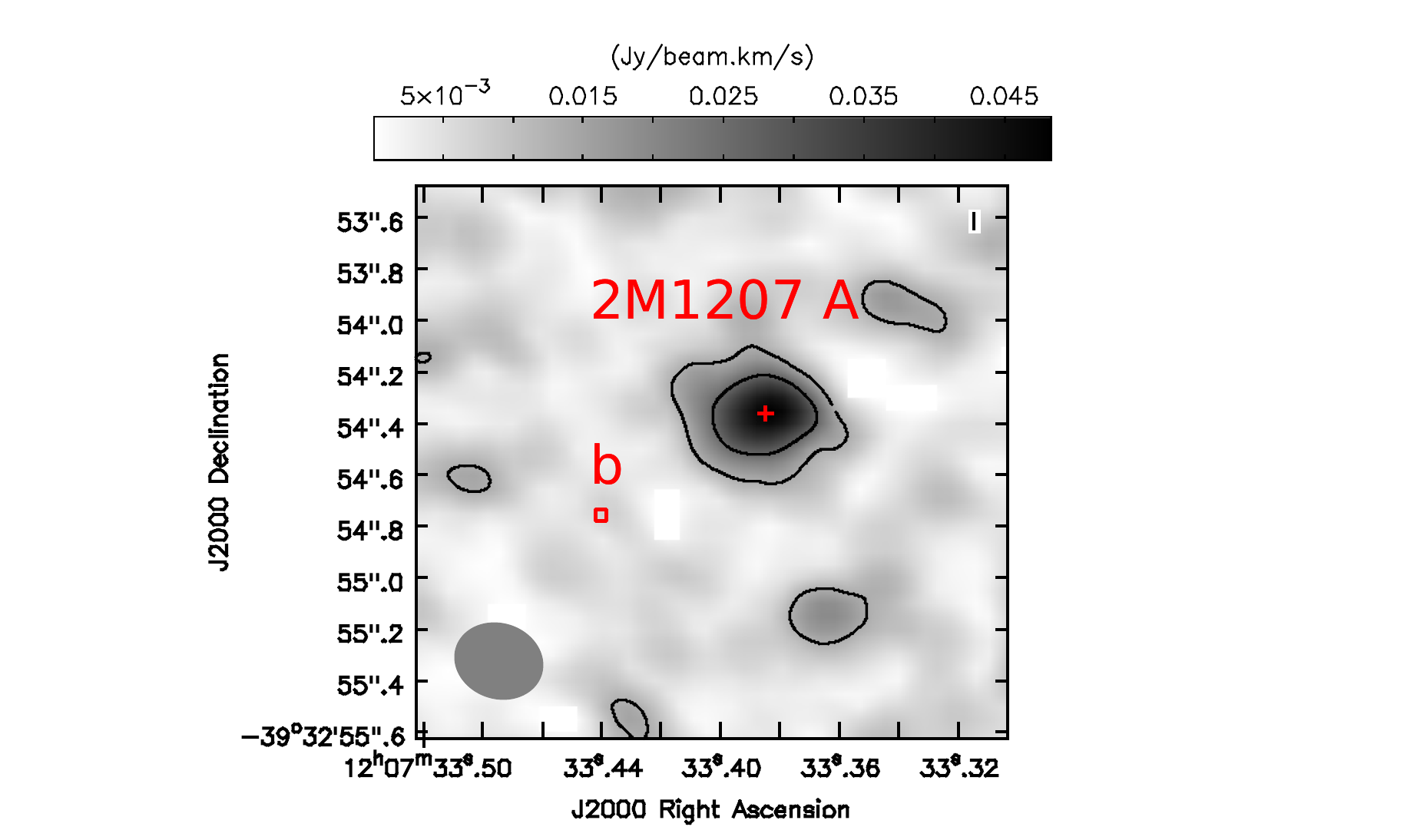}  \\

\end{tabular}
\caption{Left) ALMA Band 7 continuum map of 2M1207 at 0.89 mm. Contours are drawn at -3$\sigma$ (dashed line), 3$\sigma$, 9$\sigma$, 15$\sigma$ (solid), where 1$\sigma =$ 26 $\mu$Jy beam$^{-1}$ is the rms noise of the map. The magenta ellipse in the lower left corner represents the synthesized beam with FWHM size of $0.42'' \times 0.35''$ and a position angle of 87 deg. Right) ALMA moment 0 map of $^{12}$CO ($J = 3 - 2$) from 2M1207. Contours are drawn at 3$\sigma$, 6$\sigma$ (solid lines), where 1$\sigma =$ 4.4 mJy beam$^{-1}$ km s$^{-1}$ is the rms noise on the map. To produce this map, a clipping at about the rms noise level on each channel, i.e. about 1.8 mJy/beam, was adopted. The grey ellipse in the lower left corner represents the synthesized beam with FWHM size of $0.35'' \times 0.29''$ and a position angle of 72 deg. In each panel, the cross and square symbols indicate the location of 2M1207A and b, respectively. }
              \label{fig:2m1207}
\end{figure*}

\section{Results}
\label{sec:results}

\subsection{Continuum image}
\label{sec:continuum}

The CASA task \texttt{clean} was used to Fourier invert the complex visibilities and deconvolve the interferometric dirty image. 
The left panel on Figure~\ref{fig:2m1207} shows the ALMA map of the $\lambda$0.89 mm continuum emission from the 2M1207 system. This map was obtained using a natural weighting (Briggs robust parameter of 2) to maximize sensitivity. The angular resolution is $0.42'' \times 0.35''$ which corresponds to a spatial resolution of about 22 au $\times$ 18 au at the distance of 2M1207.

Continuum emission from the 2M1207A brown dwarf is clearly detected \citep[offsets of $\simless~0.1''$ in both right ascension and declination between the disk center and the expected 2M1207A location after considering a proper motion of ($\mu_\alpha,\mu_\delta$)=(-78,-24), mas yr$^{-1}$][]{Chauvin:2005}. The flux density from the disk is $620 \pm 67$ $\mu$Jy, where the uncertainty accounts for a rms noise of 26 $\mu$Jy beam$^{-1}$ but is dominated by an uncertainty of 10$\%$ on the absolute flux scale calibration.
The peak flux density is consistent with the total flux measured from the source, and a gaussian fit on the image plane using the CASA task \texttt{imfit} returns a point-like source convolved with the synthesized beam. This indicates that the emission is spatially unresolved, implying that the bulk of the emission comes from a region smaller than $\approx 20$ au in diameter, or 10 au in radius. 
 
Dust continuum emission is not seen at the location of the 2M1207b planetary-mass object (shown by a square symbol on Figure~\ref{fig:2m1207}). The $3\sigma$ upper limit for the flux density at 0.89 mm is 78~$\mu$Jy, assuming point-like emission.

\subsection{CO J = 3 - 2}

\begin{figure}
\centering
\includegraphics[scale=0.45,trim=0 0 0 0]{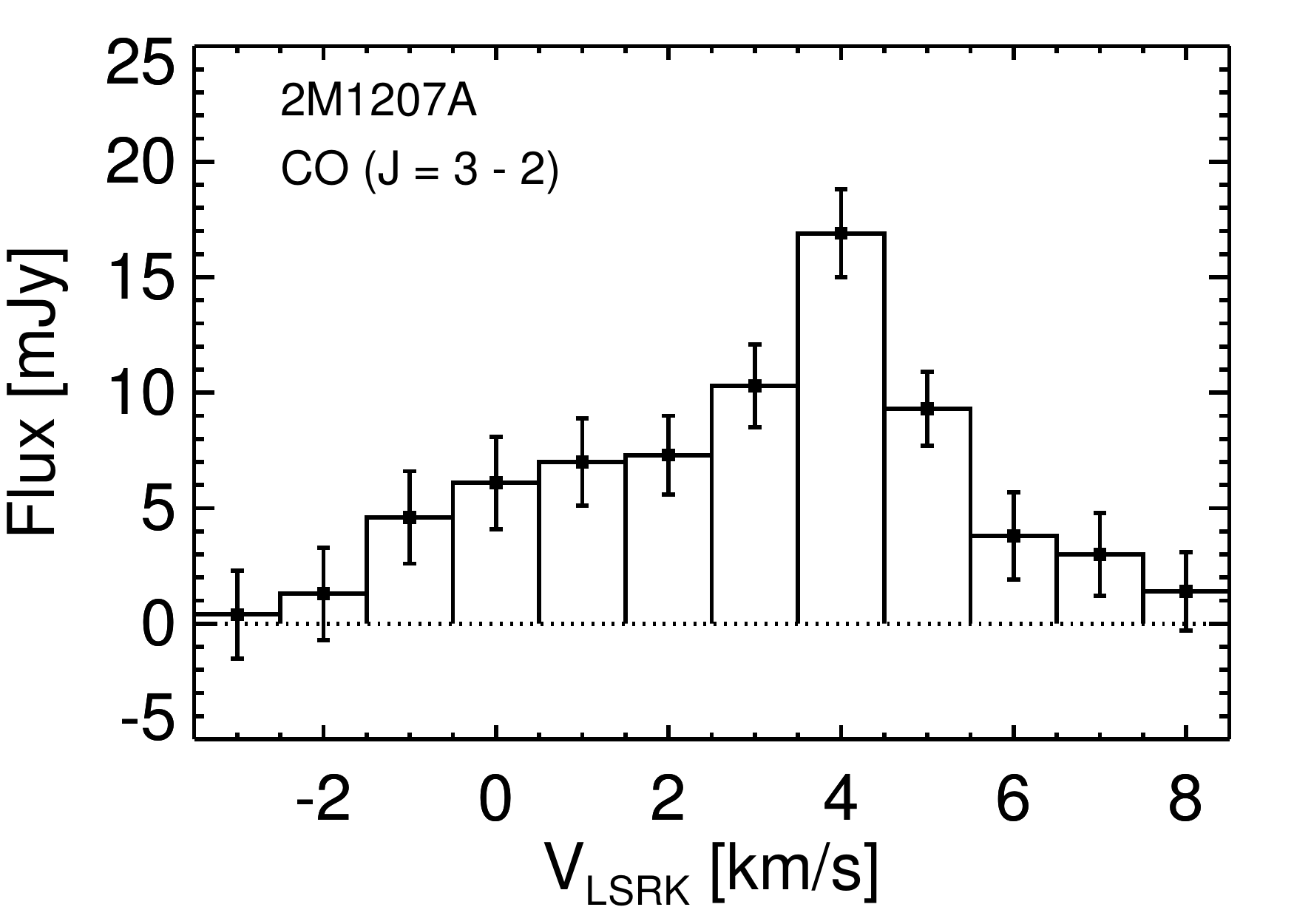} 
\caption{Spatially integrated CO($J = 3 - 2$) spectra for the 2M1207A disk. The spectrum was obtained by integrating over the pixels within a circular aperture with a radius of $0.4''$ from the disk center. The width of each bar corresponds to the velocity resolution of the observations. 
}
\label{fig:co_spectrum}
\end{figure}

The right panel on Figure~\ref{fig:2m1207} shows the moment 0 map of the rotational transition $^{12}$CO ($J = 3 - 2$). This was obtained with a Briggs robust parameter of 0.5 for the \texttt{clean} algorithm (note however that the analysis of the CO emission presented in Section~\ref{sec:analysis} was performed in the visibilities space).
As in the case of dust continuum emission, molecular emission from the $^{12}$CO ($J = 3 - 2$) line was detected from the 2M1207A brown dwarf but not from the 2M1207b planetary-mass object. The 2M1207A disk has a velocity-integrated total flux of $71.0 \pm 8.4$ mJy km s$^{-1}$. CO emission is detected at velocities ranging between $V_{\rm{LSRK}} \approx 0$ and 5 km s$^{-1}$ (Figure~\ref{fig:co_spectrum}).


\section{Analysis of the dust continuum emission}
\label{sec:continuum_analysis}

Thanks to the high sensitivity of the ALMA observations and relative proximity of the 2M1207 system, our data provide very tight constraints to the material surrounding a young planetary-mass companion. 
In the case of optically thin emission, an estimate for the dust mass $M_{\rm{dust}}$ can be derived from the measured continuum flux density $F_{\nu}$ using:

\begin{equation}  \label{eq:dust_mass}
 M_{\rm{dust}} = \frac{F_{\nu} d^{2}}{\kappa_{\nu}B_{\nu}(T_{\rm{dust}})},
\end{equation}
where $d$ is the distance, $\kappa_{\nu}$ the dust opacity coefficient and $B_{\nu}(T_{\rm{dust}})$ the Planck function evaluated at the \textit{characteristic temperature} of the emitting dust. In the case of young, very low-mass stars and brown dwarfs (spectral types later than M5), \citet{vanderPlas:2016} derived a relation between the (sub-)stellar luminosity and characteristic temperature of the surrounding dust $T_{\rm{vdP}} \approx 22 (L_{\star}/L_\odot)^{0.16}$ K. This provides an approximate estimate of the temperature of the dust dominating the sub-mm emission. If using the very low luminosity estimated for 2M1207b, $L_{\star,\rm{2Mb}} \approx 1.5 \times 10^{-4} L_\odot $ \citep{Skemer:2011}, this relation would return a very low dust temperature of $\sim 5$ K. However, given the much higher luminosity of 2M1207A \citep[$L_{\star,\rm{2MA}} \approx 2 \times 10^{-3} L_\odot$,][]{Mohanty:2007} and relatively small projected separation ($\approx 40$ au), the heating may be significantly affected by the 2M1207A radiation field. In this case, $T_{\rm{dust}} \sim 8$ K, following the \citet{vanderPlas:2016} prescription. 

To test this estimate, we generated a representative disk model with the radiative transfer code \texttt{RADMC-3D} \citep{Dullemond:2012}. For this calculation we adopted the known sub-stellar properties of 2M1207A, disk surface density properties as in Section~\ref{sec:analysis} and disk vertical structure as in \citet{Skemer:2011}. We found radiative equilibrium temperatures of $\approx 6-7$ K at the location of 2M1207b under the assumption that its orbit lies on the plane of the 2M1207A disk (see Section~\ref{sec:discussion_2m1207a}). By repeating this calculation for the putative 2M1207b disk, we found similar temperatures at $\simgreat~1$ au from the planetary-mass object. If we add up the radiation fields of both objects, we obtain a temperature of $\approx 8$ K at radii of few au within the 2M1207b disk, and we therefore adopt this value for the characteristic temperature of the dust in this disk. 

It is worth noticing that at these low heating rates, radiation from the diffuse interstellar medium may act as a relevant extra heating source. If this is true, our temperature estimate would have to be considered as a lower limit and our corresponding upper limit for the dust mass of the 2M1207b disk inferred from the ALMA non-detection (see below) may be too high. Note also that viscous heating may potentially dominate the heating budget in small circum-planetary disks for values of the mass accretion rate $\gg 10^{-6}~M_{\rm{Jup}}$ yr$^{-1}$ \citep{Isella:2014}. However, these values are ruled out for the 2M1207 system by emission line observations in the UV, optical and NIR \citep{Scholz:2005,Whelan:2007,Herczeg:2009,France:2010}.

With a distance $d = 52.8$ pc and dust opacity $\kappa_{\nu} = 3.4$ cm$^2$ g$^{-1}$ at a frequency of 338 GHz \citep{Beckwith:1990}, the $3\sigma$ upper limit for the flux density of 2M1207b presented in Section~\ref{sec:results} provides an upper limit of $M_{\rm{dust},\rm{2Mb}} < 0.013~M_\oplus \approx 1.1~M_{\rm{Moon}}$. Under the same assumptions, i.e. distance, dust opacity and temperature, the dust mass for the detected 2M1207A disk is $M_{\rm{dust},\rm{2MA}} \approx 0.1~M_\oplus$.

This upper limit for the dust mass of the 2M1207b disk is valid only in the optically thin assumption for the dust emission.
In principle, a small disk made of dense dust could produce optically thick emission but with a very low flux density because of the small radius. In the optically thick approximation, the upper limit provides an upper limit for the radius of the 2M1207b disk of $\approx 0.5$ au. The fact that the dust continuum emission observed toward 2M1207A is not spatially resolved indicates that the emitting dust in the 2M1207A disk is located mostly within a radius of $\simless~10$ au. More sensitive observations are required to probe the possible presence of very low-density dust at outer radii.

\section{Analysis of the CO emission of the 2M1207A disk}
\label{sec:analysis}

We use the \texttt{DiskJockey} package\footnote{Freely available at \url{https://github.com/iancze/DiskJockey} under an open-source MIT license.} \citep{Czekala:2015a} to model the continuum-subtracted CO emission of the 2M1207A disk. The disk structure is modeled with a parametric description of the gas density and temperature. The velocity field of the disk is dominated by the central sub-stellar mass, and so by modeling the disk structure and kinematics, we can infer the central sub-stellar mass based upon the morphology of the molecular line emission. Raw channel maps of the disk are synthesized using \texttt{RADMC-3D} \citep{Dullemond:2012}, Fourier transformed, and then sampled in the $u-v$ plane at the baselines corresponding to the ALMA observations, where their goodness of fit is evaluated using a $\chi^2$ statistic incorporating the visibility weights. Fitting directly in the $u-v$ plane ensures that we preserve the noise properties of the dataset, and are able to derive accurate exploration of the posterior probability distribution, including realistic estimates of the parameter uncertainties. Further description of the modeling framework can be found in \citet{Czekala:2015a}. The list and description of the model parameters are presented in Table~\ref{table:parameters}.

 \begin{figure*}[ht!]
\begin{center}
  \includegraphics{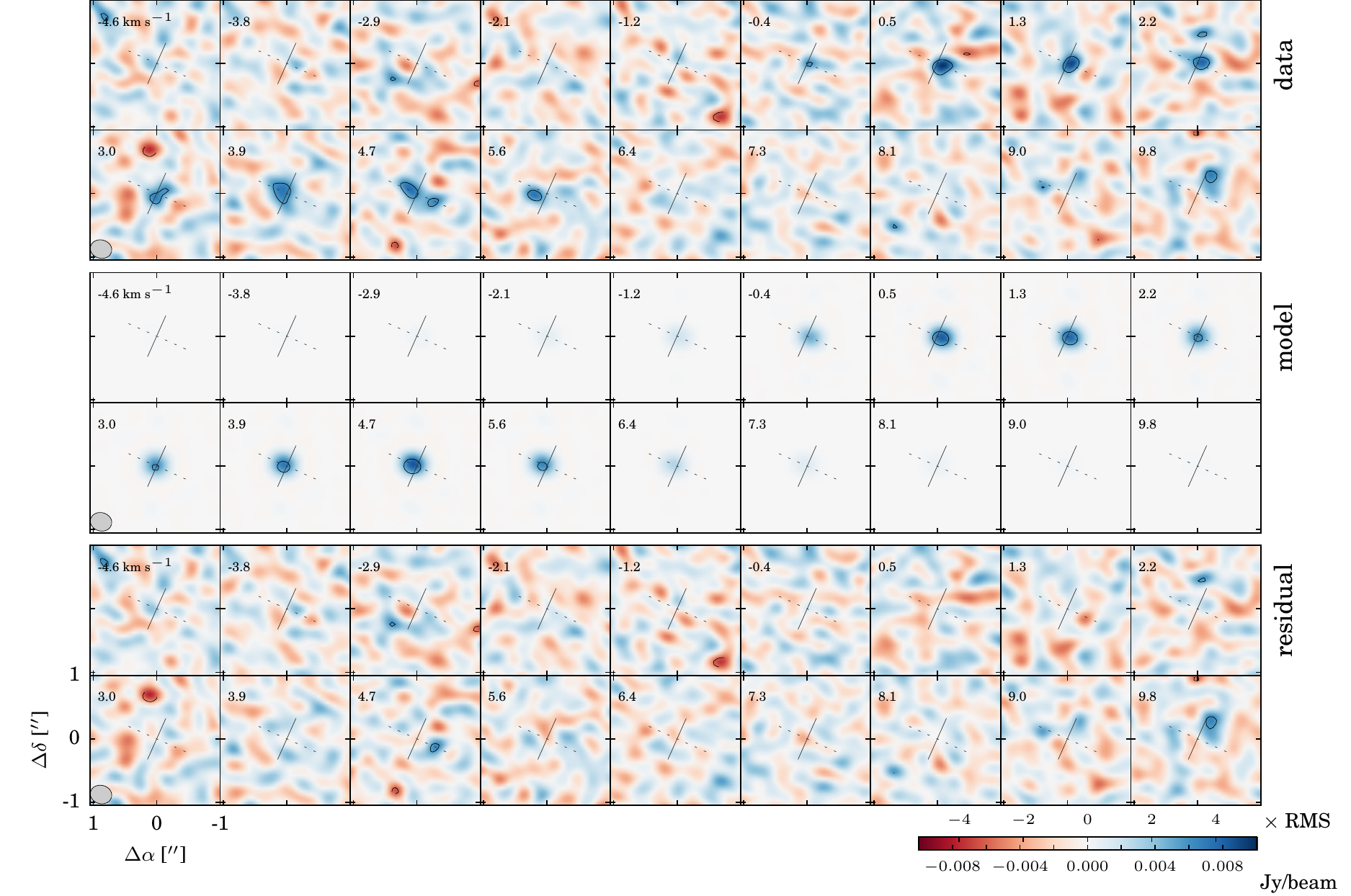}
  \figcaption{
  Data (top), best-fit model (middle) and residual (bottom) channel maps with contours drawn at multiples of 3 times the rms measured on the maps, where the rms in each channel is about 1.8 mJy/beam. Negative residuals are denoted by dashed contours. The velocity correspondent to each channel is labelled on the top left corner of each map. The synthesized beam size is shown in the left bottom corner in the bottom left channels. 
  \label{fig:chmaps}}
  \vspace{5mm}
  \end{center}
\end{figure*}

Motivated by the likely outer truncation of the 2M1207A disk by the planetary-mass companion (see Section~\ref{sec:discussion_2m1207a}), we explore models for the disk surface density:
\begin{equation}
\Sigma(r) = \Sigma_c \left (\frac{r}{r_c} \right)^{- \gamma} \exp \left[ - \left(\frac{r}{r_c} \right)^{2 - \gamma_e} \right],
\end{equation}
where the gradient $\gamma_e$ of the outer exponential taper is allowed to float separately from the power law surface density gradient $\gamma$ and can deliver disk models that have a sharply truncated outer radius. For these models, we assume a prior for $\gamma_e \leq 2$, which includes the case of a disk with a pure power-law surface density profile with no exponential taper ($\gamma_e = 2$). However, our analysis does not provide significant constraints on this parameter.

The limited signal-to-noise ratio of the CO data does not allow us to constrain all the structural parameters for these disk models independently.  
We explore surface density profiles with a power law exponent of $\gamma = 1$. We also include a prior that the gradient of the power-law describing the temperature profile $q$ must be between 0 and $3/4$, the maximum value allowed for a passively irradiated disk. 

\begin{deluxetable*}{llr}  
  \tablecaption{\label{table:parameters}Inferred Parameters for 2M1207A}
  \tablehead{\colhead{{\sc Parameter}} & {\sc Description} & {\sc Value}}
  \startdata
  $M_\ast$           &    Central object mass    ($M_\odot$) & 0.06$^{+0.08}_{-0.02}$  \\ 
  $i_d$       & Disk inclination (${}^\circ$) & 35$^{+20}_{-15}$ \\
  P.A. & Disk position angle\tablenotemark{a}       (${}^\circ$) & 174 $\pm$ 12  \\
  $r_c$                 &    Characteristic radius    (au) & 9.4 $\pm$ 1.5 \\  
  $\log (\Sigma_{c})$ & Scaling factor for the surface density (log g cm$^{-2}$)  & -3.55 $\pm$ 0.35  \\
  $\gamma_e$      & Radial gradient of the outer exponential taper        &  $\leq 2$ (unconstrained) \\
  $T_{10}$ &  Temperature at 10 au (K) & 0 $-$ 300 (unconstrained)   \\
  $q$          &  Temperature power-law index                     & 0 $-$ 0.75 (unconstrained) \\ 
  $v_{\rm sys}$ &  Systemic velocity\tablenotemark{b}     ($\textrm{km s}^{-1}$) & 2.72 $\pm$ 0.19  \\
  $\xi$                  &    Non-thermal broadening line width   ($\textrm{km s}^{-1}$) & 0.67 $\pm$ 0.40  \\
  $\delta_\alpha$ &    RA offset  (\arcsec) & -0.94 $\pm$ 0.02  \\
  $\delta_\delta$  &    DEC offset    (\arcsec) & -0.36 $\pm$ 0.01  \vspace{1mm}
  \enddata
  \tablecomments{The quoted best-fit values correspond to the peaks of the marginal posterior distributions.  The uncertainties correspond to the 68.3\%\  confidence intervals.}
  \tablenotetext{a}{The Position Angle is measured as the angle from North to the projection of the disk angular momentum vector in the sky plane (North towards East).}  
  \tablenotetext{b}{In the LSRK frame, for the standard radio definition.}  
  
\end{deluxetable*}

Because we are modeling angular separations on the sky, the dynamical modeling does not provide any information about the distance to the source on its own. Therefore, we proceed to fit the source using a distance prior of $d = 52.8 \pm 1.0\,\textrm{pc}$, determined using a weighted mean of recent parallaxes compiled by E. Mamajek~\citep{Gizis:2007,Biller:2007,Ducourant:2008}, with the most accurate contribution from \citet{Ducourant:2008}\footnote{See \url{www.pas.rochester.edu/$\sim$emamajek/memo\_2m1207.html}}.
The uncertainties in the model parameters originating from the distance prior are small compared to the statistical uncertainties from the dynamical modeling.

We explore the posterior distribution of parameters using the Markov Chain Monte Carlo (MCMC) ensemble sampler \citep{Goodman:2010,Foreman:2013}, with 48 walkers run for 10,000 iterations. After the first 5,000 iterations were discarded for burn-in, we computed the autocorrelation time to be 80 iterations, ensuring that we have sufficient independent samples from the posterior. 

\begin{figure}[t!]
\begin{center}
  \includegraphics[scale=1.5]{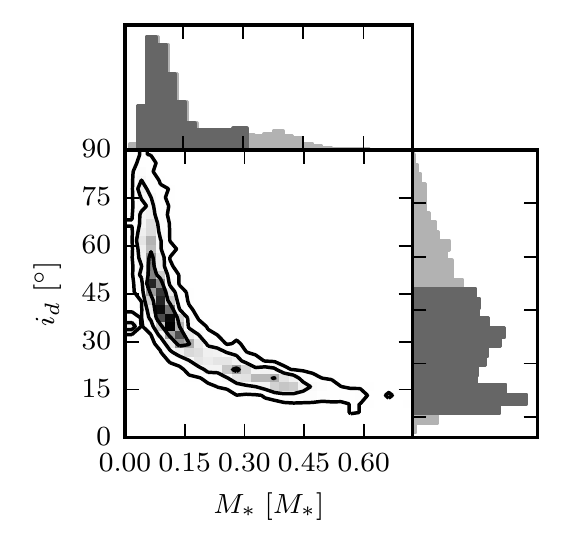}
  \figcaption{
  The joint $\{ M_\ast,\, i_d \}$ posterior with 1, 2, and 3 sigma contours. On either side are shown the marginalized 1-dimensional posteriors with the highest density interval containing 68.3\% of the samples shaded in dark gray.
  \label{fig:posterior}}
  \end{center}
\end{figure}

The best-fit parameters and 68.3\% confidence intervals are described in Table~\ref{table:parameters}, and the joint $\{M_\ast,\,  i_d \}$ posterior shown in Figure~\ref{fig:posterior}. We demonstrate the quality of this fit to reproduce the CO emission by showing the data, model, and residual channel maps in Figure~\ref{fig:chmaps}. 

Given the inferred disk position angle of $174^{\circ} \pm 12^{\circ}$ the disk is consistent with being perpendicular to the observed optical outflow \citep[measured position angle of $65^{\circ} \pm 10^{\circ}$,][]{Whelan:2012}. 
We infer a dynamical estimate for the mass of 2M1207A of $M_\ast = 60^{+80}_{-20}\,M_{\rm{Jup}}$. This is marginally consistent with the results by \citet{Gizis:2002,Mamajek:2005,Mohanty:2007,Skemer:2011}, all of whom found $\sim 25~M_{\rm{Jup}}$ from the Lyon theoretical models to reproduce the effective temperature inferred from near-IR spectra and the age estimated assuming 
membership of the TW Hydrae association. The uncertainty of these mass estimates is typically reported as $\approx 3~M_{\rm{Jup}}$ and is dominated by the assumptions on the formation pathway and accretion history \citep[see discussions in][Section 3]{Mamajek:2005,Bowler:2016}.

The ALMA CO observations spatially resolve the 2M1207A disk only marginally. This results in a significant degeneracy along the product $M_\ast \sin^2 i_d$, where more face-on disk inclinations yield higher mass estimates. As shown in Figure~\ref{fig:posterior}, the ALMA data alone cannot rule out stellar-like masses ($\simgreat~0.08~M_\ast$) for disk inclinations ${15}^\circ \simless i_d \simless {30}^\circ$. Lower masses, below the hydrogen burning limit, are obtained if one considers the range of higher inclinations, ${70}^\circ \simless i_d \simless {75}^\circ$, favored by the SED-fitting analysis performed by \citet{Skemer:2011} using \texttt{RADMC} flared disk models.
Under the assumptions presented above, our analysis derives a value of $9.4 \pm 1.5$ au for the characteristic disk radius. 
The discussion of this result is presented in Section~\ref{sec:discussion_2m1207a}. 

\section{Discussion and Conclusion}
\label{sec:discussion}

We discuss here the main results of our analysis on the ALMA data of the 2M1207 system. In Section~\ref{sec:circum-planetary} we discuss the upper limits obtained so far for the mass of dust surrounding young planetary-mass companions, whereas Section~\ref{sec:discussion_2m1207a} focuses on the properties of the 2M1207 system.

\subsection{Upper limits on the mass of dust orbiting young planetary-mass companions}
\label{sec:circum-planetary}

\begin{figure}[t!]
\begin{center}
  \includegraphics[scale=0.5]{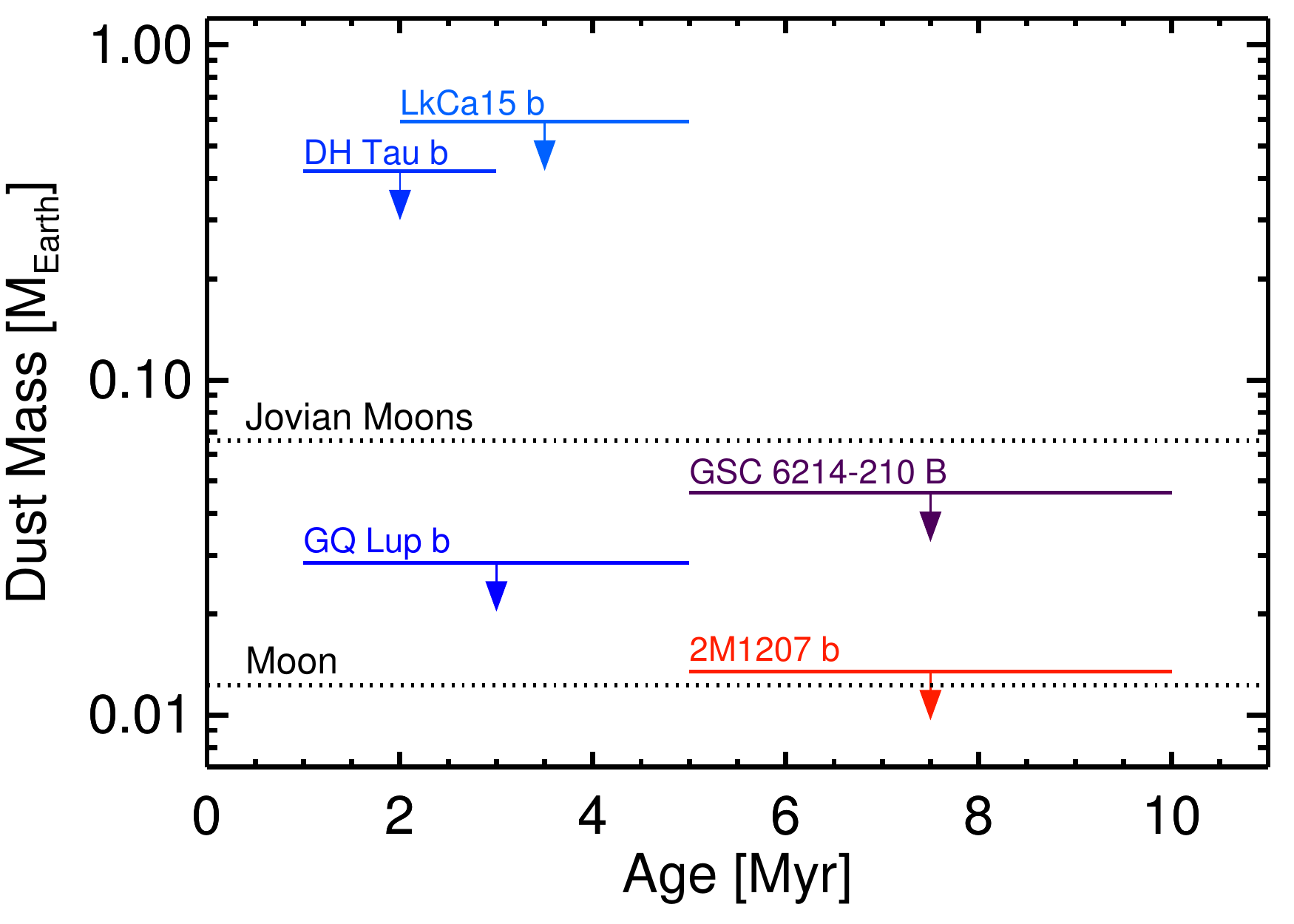}
  \figcaption{
  Dust mass vs age for material surrounding four young companions with estimated masses close to or below the deuterium burning limit. The upper limit for 2M1207 b was derived in this work (Section~\ref{sec:continuum_analysis}), whereas for the other objects we converted the measured upper limits on the sub-mm/mm fluxes assuming optically thin dust emission, the \citet{Beckwith:1990} opacity law, and the \citet{Andrews:2013} prescription for the dust temperature (see text).  
  \label{fig:circum-planetary}}
  \end{center}
\end{figure}

The upper limit of $\sim 1.1~M_{\rm{Moon}}$ derived for the dust mass of the 2M1207b disk is lower than for any other young companion with an estimated mass close to or below the deuterium burning limit ($\approx 13~M_{\rm{Jup}}$). Figure~\ref{fig:circum-planetary} shows the upper limits derived for the dust mass (in the optically thin assumption) of any circum-planetary material together with the estimated ages for the host stars or brown dwarfs. These include the. LkCa15 b candidate planetary-mass companion \citep[$\sim 6 - 10~M_{\rm{Jup}}$,][]{Kraus:2012}, which was targeted by high-angular resolution observations with the Very Large Array (VLA) at $\sim 7$ mm \citep{Isella:2014}, GSC 6214-210 B ($\sim 15~M_{\rm{Jup}}$) and GQ Lup b ($10 - 36~M_{\rm{Jup}}$) observed with ALMA at $0.87$ mm \citep[][respectively]{Bowler:2015,MacGregor:2016}, DH Tau b ($8 - 21~M_{\rm{Jup}}$), recently observed at $1.3$ mm with the Northern Extended Millimeter Array \citep[NOEMA, ][]{Wolff:2017}. The upper limits for the dust mass of the putative circum-planetary disks shown in Figure~\ref{fig:circum-planetary} were derived from the upper limits for the sub-mm/mm flux densities in the literature using Eq.~\ref{eq:dust_mass}, which assumes optically thin emission. We adopted the opacity law from \citet{Beckwith:1990} and the dust temperature using the prescription from \citet{Andrews:2013} which calculates a characteristic dust temperature in the outer regions of a disk surrounding a star of a given luminosity. This approach gives the most conservative upper limit for the dust mass \citep[a higher $T_{\rm{dust}}$, as might be expected for a disk smaller than the 200 au assumed in Andrews et al., would result in a lower dust mass upper limit, see][] {Hendler:2017}.
We do not attempt a more precise calculation of the dust temperature for each disk, e.g. via radiative transfer calculations, because the uncertainty on the dust mass estimate is dominated by the uncertainty on the dust opacity coefficient $\kappa_{\nu}$. 

We did not include the FW Tau system, as the nature of the FW Tau C companion with a detected disk with dust mass of $\approx 1 - 2~M_\oplus$ \citep{Kraus:2015} is very unclear: its SED can be reproduced by a planetary-mass companion as well as by a very low mass star or brown dwarf \citep{Bowler:2014}. 

Although it may be misleading to compare constraints for systems with very different properties (mass of the host star/brown dwarf, host-companion physical separation, companion mass, disk properties) and possibly also different formation mechanisms, these first results from very sensitive sub-mm observations of young planetary-mass companions indicate that very little material in the form of mm-sized pebbles is present around these objects. 
This seems to be in apparent contrast with several models of planet formation in which relatively massive circum-planetary disks or envelopes are expected to feed the proto-planets.  

For example, \citet{Stamatellos:2015} predict circum-planetary disks to retain a gas mass of $\sim 10^{-2} - 1~M_{\rm{Jup}}$ for longer than 10 Myr (thus including the ages of all the observed systems) around $\sim 10 - 30~M_{\rm{Jup}}$ objects formed via the fragmentation of gravitationally unstable disks. In order to be consistent with the dust mass upper limits for companions of similar masses, the dust-to-gas mass ratio would have to be lower than the ISM-value of $10^{-2}$ by up to two orders of magnitude. 
\citet{Zhu:2016} have calculated the sub-mm emission of circum-planetary disks accounting for shock-driven accretion and found that ALMA should be able to detect circum-planetary disks with different accretion properties. The observational limits rule out several of these models. 
In particular, ``minimum mass sub-nebula'' models, of the kind proposed to form the satellites of Jupiter and Saturn in the Solar System \citep{Pollack:1984}, require a minimum disk mass of $\sim 0.02$ times the planet mass. These models would produce flux densities at sub-mm wavelengths well above the ALMA sensitivity of the aforementioned observations. Instead, the current observational limits are consistent with the ``gas-starved'' models by \citet{Canup:2002,Canup:2006}, who predicted significantly lower densities for the circum-planetary disks around Jupiter and Saturn after accounting for the effects of satellite-disk interaction \citep{Zhu:2016}, as well as with recent hydrodynamical simulations by \citet{Szulagyi:2016}, which predict a roughly linear relation between the circum-planetary disk mass and the mass of the parental circumstellar disk, with the former being two orders of magnitude lower than the latter for both the core-accretion and gravitational instability scenarios.   

It is also possible that, at least for some of these systems, the little amount of disk material is due to outward scattering after a dynamical interaction with another massive companion closer to the star/brown dwarf. Besides investigating the presence of other massive planets (or brown dwarfs) closer to the central object, sub-mm observations with very high angular resolution and sensitivity could reveal possible signatures of this interaction imprinted in the circumstellar disk structure. Another consequence of the scattering scenario is the relatively high eccentricity expected for the scattered planets, and high-precision astrometric observations may test this prediction by measuring their proper motions \citep[see discussions in][]{Bowler:2014,MacGregor:2016}.

\subsection{Properties of the 2M1207A disk}
\label{sec:discussion_2m1207a}

Although observations with higher angular resolution and sensitivity are needed to better constrain the detailed structure of the 2M1207A disk, the ALMA observations presented in this work reveal a very compact disk. 
The analysis described in Section~\ref{sec:analysis} derives an estimate for the disk characteristic radius of $\approx 10$ au, and the fact that the dust emission was spatially unresolved points toward a radius $\simless~10$ au in dust. A more concentrated distribution of dust particles relative to gas is predicted by models of radial migration of solids in gas-rich disks \citep{Weidenschilling:1977}. However, observations with better sensitivity and angular resolution are necessary to better characterize the distribution of low-density material in the outer regions of the 2M1207A disk \citep[see][]{Hughes:2008}.

Relative to other disks orbiting young brown dwarfs and very low-mass stars, the 2M1207A disk is significantly smaller than the most massive disks ($M_{\rm{dust}} \approx 2 - 6~M_{\oplus}$) known around brown dwarfs and very low-mass stars in the younger, $\sim 1 - 3$ Myr-old Taurus region, which show outer radii $\simgreat~60$ au \citep{Ricci:2014,Testi:2016}. Upper limits of $\simless~20$ au were instead derived for the dust emission from the majority of brown dwarf disks in the younger Ophiuchus region \citep{Ricci:2012,Testi:2016}. 

In the case of the 2M1207A disk, a natural explanation for its small radius is the tidal truncation of the outer disk regions by the gravity of 2M1207b. 
If 2M1207b was formed in situ, for example via gravitational fragmentation of the 2M1207A disk as proposed by \citet{Lodato:2005}, then the 2M1207A disk would be much smaller today than when it formed 2M1207b which is seen at a projected separation of $\approx 40$ au. The later evolution of the disk would then be strongly affected by the gravity of 2M1207b itself. 
According to the theoretical models of \citet{Artymowicz:1994}, the disk should be externally truncated at a fraction of the component separation ($\sim 0.2 - 0.5a$), and this effect could therefore explain the small radial extent of the 2M1207A disk. 

However, because of the wide separation and consequent long orbital period, the orbital parameters are not known for this system and the hypothesis of a \textit{physical} separation much larger than the \textit{projected} separation cannot be ruled out. 
If we assume that the orbit of 2M1207b lies on the plane of the 2M1207A disk, then we can use our constraints on the disk inclination and position angle to derive a posterior probability distribution for the current separation. From this distribution, we derived an estimate of $42^{+19}_{-2}$ au for the current separation (uncertainties at the 68\% confidence level), with the lower limit corresponding to the projected separation.  

\begin{figure}
\centering
\includegraphics[scale=0.45,trim=0 0 0 0]{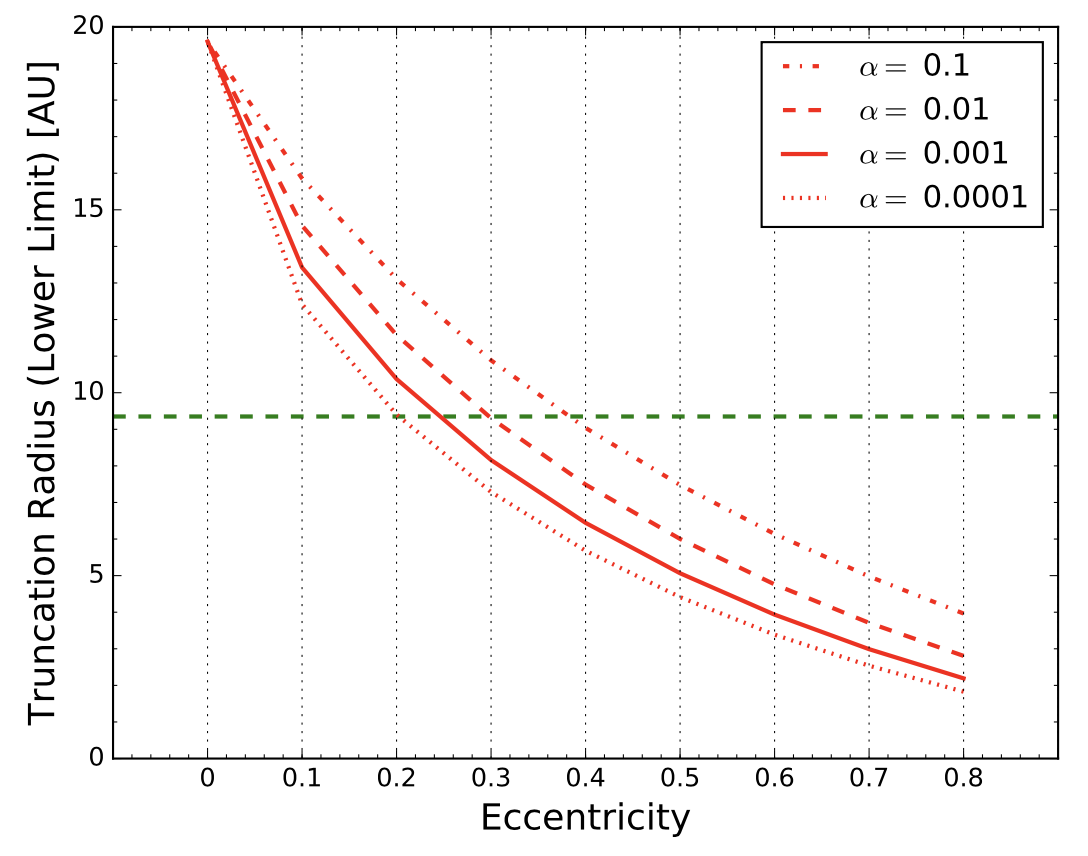} 
\caption{Lower limits for the truncation radius for different orbital eccentricities of the 2M1207 binary system and different values for the viscosity $\alpha$-parameter. The green dashed line represents the best-fit value for the characteristic radius $r_c$ for the 2M1207A disk (Section~\ref{sec:analysis}), and is shown for comparison. For this analysis we adopted masses of 60 and 5$~M_{\rm{Jup}}$ for 2M1207A and b, respectively.
}
\label{fig:truncation_analysis}
\end{figure}


In order to account for our ignorance on the true orbital parameters of the 2M1207 system and yet quantify the likelihood that tidal truncation models can reproduce the observed radius of the 2M1207A disk, we applied the Monte Carlo method presented by \citet{Harris:2012}. This was used to construct a probabilistic model for the tidal truncation radius using only the projected separation and mass ratio of the two companions. We assumed uniform prior distributions for the unknown orbital parameters, while to derive the relation between the truncation radius and orbital parameters we fit the results from the \citet{Artymowicz:1994} tidal truncation models for different values of the viscosity $\alpha$-parameter chosen between 0.1 and 0.0001. 

The inferred relation between the lower limit of the disk truncation radius and binary eccentricity and viscosity $\alpha$-parameter is shown in Figure~\ref{fig:truncation_analysis}.
As seen in the figure, values of $\simgreat~10$ au for the tidal truncation radius, consistent with the observations of the 2M1207A disk, 
are found for eccentricities $e \simless 0.4$ (this range of eccentricity values becomes $e \simless 0.3$ if we assume that the orbit of 2M1207b lies on the plane of the 2M1207A disk). Larger eccentricity values give predominantly lower values for the tidal truncation radii as they produce more orbits with shorter periastra. 
This shows that, under the reasonable assumptions presented here, the tidal truncation models can naturally explain the small radius of the 2M1207A disk given the observational information currently available for this system.  

Although the results of this analysis suggest that the evolution of the 2M1207A disk has been likely affected by tidal truncation from 2M1207b, it is worth noting that both the dust mass inferred for the 2M1207A disk and the upper limit for 2M1207b follow the scaling relation between dust mass and stellar/sub-stellar mass, i.e. $\log (M_{\rm{dust}}/M_{\oplus}) = (1.9 \pm 0.4) \times \log (M_{\star}/M_{\odot}) +  (0.8 \pm 0.2)$, found for more massive brown dwarfs and pre-Main Sequence stars in the $\sim 5-10$ Myr old Upper Sco region~\citep[][]{Barenfeld:2016,Pascucci:2016}. This behaviour is expected if the two objects in the 2M1207 binary system were born in two independent protostellar cores, rather than from the same disk. The observational data obtained so far cannot rule out any of these two competing mechanisms for the formation of the 2M1207 system.

\section{Conclusions}
\label{sec:conclusions}

We presented new ALMA observations for the dust continuum emission at 0.89 mm and CO J=3-2 line emission for the young substellar binary system 2M1207. The main results are as follows.

\begin{itemize}

\item The disk around the brown dwarf 2M1207A was detected in both dust continuum and CO J=3-2 line emission. The dust emission was spatially unresolved at the angular resolution of our ALMA observations. This indicates that the bulk of the dust emission comes from disk radii $\simless~10$ au.

\item Neither dust emission nor CO J=3-2 line emission were detected at the location of the planetary-mass companion 2M1207b. Under the assumption of optically thin dust emission, we estimated a 3$\sigma$ upper limit of $\sim 1~M_{\rm{Moon}}$, which is the tightest upper limit obtained so far for the mass of dust particles surrounding a young planetary-mass companion. In the optically thick limit, our ALMA non-detection translates into a 3$\sigma$ upper limit of $\approx 0.5$ au for the 2M1207b disk radius.

\item We fit the channel-dependent interferometric visibilities measured for the CO J=3-2 line emission for the 2M1207A disk using the \texttt{DiskJockey} package. Adopting a power-law function with an exponential taper for the radial dependence of the gas surface density, our analysis infers a disk characteristic radius of $9.4 \pm 1.5$ au. The disk inclination and position angle are $35^{+20}_{-15}$ degrees  and $174 \pm 12$ degrees, respectively. We also obtained a dynamical estimate of $60^{+80}_{-20}~M_{\rm{Jup}}$ for the mass of 2M1207A.

\item The small size of the 2M1207A disk is likely due to the effect of tidal truncation by 2M1207b. If 2M1207b lies on the plane of the 2M1207A disk, the current physical separation between the two companions is $42^{+19}_{-2}$ au (68$\%$ confidence level).

\end{itemize}

Future ALMA observations with better sensitivity and higher angular resolution than the ones presented in this work will provide more stringent constraints for the dynamical mass of 2M1207A as well as for the physical structure of its disk.




\acknowledgments

This paper makes use of the following ALMA data: ADS/JAO.ALMA\#2013.1.01016.S. ALMA is a partnership of ESO (representing its member states), NSF (USA) and NINS (Japan), together with NRC (Canada) and NSC and ASIAA (Taiwan), in cooperation with the Republic of Chile. The Joint ALMA Observatory is operated by ESO, AUI/NRAO and NAOJ.
The National Radio Astronomy Observatory is a facility of the National Science Foundation operated under cooperative agreement by Associated Universities, Inc. JMC acknowledges support from the National Aeronautics and Space Administration under Grant No. 15XRP15\_20140 issued through the Exoplanets Research Program.
Support for this work was provided by NASA through Hubble Fellowship grant HST-HF2-51369.001-A awarded by the Space Telescope Science Institute, which is operated by the Association of Universities for Research in Astronomy, Inc., for NASA, under contract NAS5-26555.

\end{document}